\documentclass[11pt]{article}

\usepackage[a4paper,margin=2.5cm]{geometry}
\usepackage{amsmath,amssymb,amsthm}
\usepackage{graphicx}
\usepackage{hyperref}
\usepackage{comment}
\usepackage{braket}
\usepackage{bm}

\newcommand{\Sb}{\bar{S}}

\title{Symmetry constraints on magnon-induced dynamics of skyrmion helicity in frustrated magnets}

\author{
Gabriel Mayor$^{1}$ and Christina Psaroudaki$^{1}$%
\thanks{Corresponding author: \texttt{christina.psaroudaki@phys.ens.fr}}
\\[0.5em]
\small $^{1}$Laboratoire de Physique de l'École Normale Supérieure, ENS, Université PSL, CNRS,\\
\small Sorbonne Université, Université de Paris, F-75005 Paris, France
}

\date{}

\begin{document}

\maketitle

\begin{abstract}
The helicity of a magnetic skyrmion in a centrosymmetric frustrated magnet is a soft collective coordinate and a candidate variable for encoding quantum information. Its coherence is ultimately influenced by coupling to the magnon environment, which is commonly incorporated only phenomenologically. Here we combine a microscopic Keldysh treatment of the helicity--magnon coupling to resolve the contribution of individual magnon modes to the effective helicity dynamics. In the \(U(1)\)-symmetric model, the magnon bath generates, to leading order, a reactive helicity mass renormalization. When \(U(1)\) symmetry is broken, additional direct helicity--magnon couplings become allowed. In the presence of finite magnon lifetimes, these couplings generate an Ohmic contribution to the low-frequency helicity dynamics. The spatial symmetry of the perturbation determines which magnon sectors contribute to this damping, providing mode-resolved selection rules for the helicity--magnon coupling. Thus, we establish a microscopic connection between helicity dynamics and the excitation spectrum of frustrated skyrmions and identify the symmetry constraints governing dissipation of helicity-based quantum states.
\end{abstract}

\section{Introduction}

The dynamics of collective degrees of freedom embedded in an environment of microscopic excitations is a central problem in nonequilibrium condensed-matter physics\cite{Weiss2012}. Integrating out the environmental modes generally produces both fluctuating forces and nonlocal-in-time response kernels, so that the resulting dynamics can differ qualitatively from the local damping assumed in phenomenological descriptions \cite{LindenbergWest1990}. This distinction becomes particularly important at low temperatures, where the quantum nature of the bath cannot be neglected\cite{PhysRevLett.116.225301,PhysRevA.95.053604}. For magnetic textures, a natural source of such intrinsic environmental fluctuations is provided by the spin-wave excitations around the classical magnetic configuration\cite{PhysRevB.53.3237,PhysRevX.7.041045}.

Magnetic skyrmions provide a particularly rich setting in which to study this problem. Their low-energy dynamics can be described in terms of a small number of collective coordinates coupled to an extended spectrum of magnon excitations \cite{PhysRevLett.100.127204,PhysRevB.95.174408,PhysRevB.99.054430}. Conventionally, dissipation in skyrmion dynamics is introduced through the Gilbert term of the Landau--Lifshitz--Gilbert equation, corresponding to a local, Ohmic friction with a phenomenological damping parameter\cite{Gilbert2004}. A microscopic treatment, however, reveals a considerably richer structure. For the skyrmion center-of-mass, integrating out quantum magnon fluctuations generates a frequency-dependent, non-Markovian damping kernel and colored stochastic forces related through the fluctuation--dissipation theorem\cite{PhysRevB.100.134404}. In the low-frequency regime, the equilibrium magnon bath produces a reactive contribution that can be interpreted as a quantum correction to the skyrmion mass, while driving the magnon bath out of equilibrium can generate additional friction-like terms absent in the static limit \cite{PhysRevLett.120.237203}. 

Magnetic skyrmions possess internal collective degrees of freedom beyond their translational motion\cite{PhysRevB.99.054430}. In centrosymmetric frustrated magnets, the helicity is a continuous $U(1)$ degree of freedom describing the global orientation of the in-plane spins. In the quantum regime, helicity can become quantized and exhibit tunnelling between distinct states, motivating proposals for helicity-based skyrmion qubits\cite{PhysRevLett.127.067201,PhysRevB.106.104422,Psaroudaki2026PhysicsToday}. The possibility of exploiting helicity quantum mechanically makes its coupling to the surrounding magnetic degrees of freedom particularly important. Understanding how these excitations modify the helicity dynamics is therefore essential for determining its effective inertia, dissipation, and ultimately the robustness of helicity-based quantum states. Recent magnon-spectrum calculations reveal strong hybridization with internal skyrmion dynamics, particularly breathing modes coupled to helicity precession, making the magnon eigenbasis the natural microscopic description of this coupling \cite{Hullahalli2026}.

In this work, we develop a microscopic theory of the dynamics of skyrmion helicity coupled to magnon excitations. We separate the helicity and its canonically conjugate longitudinal spin from the remaining spin-wave fluctuations and integrate out the latter using the real-time Keldysh formalism. This procedure first leads to non-Markovian response kernels in the effective helicity dynamics, reflecting the memory effects induced by the magnon bath. In the $U(1)$-symmetric limit, the coupling to magnons produces a reactive correction to the helicity inertia, leading to a renormalization of its effective mass. When the symmetry is broken, additional direct helicity--magnon couplings become allowed and generate mixed response kernels, including a contribution linear in frequency in the low-frequency helicity dynamics. This distinction is particularly important in the quantum regime, where the emergence of an Ohmic contribution can directly limit helicity coherence, in contrast to the weaker low-frequency effects associated with super-Ohmic coupling \cite{RevModPhys.59.1}.

The paper is organized as follows. Section~\ref{sec:model} introduces the lattice model, the continuum limit, and the separation of the helicity sector from the magnon fluctuations. Section~\ref{sec:keldysh} summarizes the Keldysh derivation of the effective helicity action and states the two central results: the mass renormalization in the symmetric model and the Ohmic coefficient in the presence of U(1) breaking. Section~\ref{sec:spectrum} maps the abstract kernels onto the computed magnon spectrum, establishing the selection rules. Section~\ref{sec:estimates} discusses implications for helicity-qubit coherence, and Sec.~\ref{sec:conclusion} concludes.

\section{Model and collective coordinates}
\label{sec:model}

We consider classical spins $\mathbf{S}_i$ on a two-dimensional triangular lattice,
\begin{align}
H = -\frac{\bar J_1}{2}\sum_{\langle i,j\rangle} \mathbf{S}_i\cdot\mathbf{S}_j
 + \frac{\bar J_2}{2}\sum_{\langle\langle i,j\rangle\rangle} \mathbf{S}_i\cdot\mathbf{S}_j
 - \bar B_z\sum_i S_i^z - \bar K \sum_i (S_i^z)^2 ,
\label{eq:Hlattice}
\end{align}
where $\bar{J}_1,\bar{J}_2>0$ denote ferromagnetic nearest-neighbor and competing antiferromagnetic next-nearest-neighbor exchange interactions, respectively, while $\bar{B}_z$ is an external magnetic field along $\hat{\mathbf z}$ and $\bar{K}$ is an easy-axis anisotropy. The Hamiltonian is centrosymmetric and invariant under global rotations about the $z$ axis, i.e., it possesses a U(1) symmetry in spin space with conserved generator $\mathcal{S}^z=\sum_i (S-S^z_i)$. The zero-temperature phase diagram hosts a fully polarized state, a skyrmion crystal, and metastable isolated skyrmions above the critical field $\bar{B}_z=\bar{J}_2-2\bar{K}$ \cite{Leonov2015,PhysRevB.93.184413,Hullahalli2026}. In the continuum limit $\mathbf{S}_i = S\mathbf{m}(\mathbf{r},t)$, the free-energy density reads
\begin{align}
\mathcal{F} = -\frac{J_1}{2}(\nabla\mathbf{m})^2 + \frac{J_2 a^2}{2}(\nabla^2\mathbf{m})^2 - \frac{H}{a^2}m_z - \frac{K}{a^2}m_z^2 ,
\label{eq:freeenergy}
\end{align}
where $J_1= \sqrt{3}\,S^2\,(3\bar J_2-\bar J_1)$, $J_2 = \sqrt S^2\,(9\bar J_2-\bar J_1)/16$, 
$H=2\,S\,\bar B_z/\sqrt{3}$, and $K=2 S^2\,\bar K/\sqrt{3}$. 
A single skyrmion above the ferromagnetic backgound is described in spherical angles $\mathbf{m}=(\cos \Phi \sin\Theta, \sin \Phi \sin \Theta,\cos \Theta)$ by $\Theta(\mathbf{r})=\Theta_0(\rho)$ and $\Phi = \Phi_0+\phi=-Q\varphi + \phi$, with $\varphi$ the polar angle around the skyrmion center, $Q$ the topological charge, and $\phi(t)$ the helicity. Because the energy is independent of $\phi$, the helicity is an exact zero mode. Its canonically conjugate momentum is the total spin deviation
\begin{align}
S_z = \bar{S} \int d^2r\, \big(1-\cos\Theta\big),
\end{align}
which is conserved in the symmetric model. The real-time coherent-state action $\mathcal{S}$ for a thin magnetic insulator in physical units of space $\mathbf{r}$ and time $t$ is given by 
$\mathcal{S}=\int d^2r dt[S\dot{\Phi}(\cos\Theta-1)-\mathcal{F}]$ \cite{Fradkin2013}. Quantization of the path integral variables necessitates the introduction of quantum fluctuations $\xi,\zeta$ around the skyrmion saddle point, which are defined in a rotating frame, with orthogonality constraints that remove the zero mode from the fluctuation space \cite{PhysRevX.7.041045,5c9q-nsb4}. The real-time action then takes the canonical form,
\begin{align}
\mathcal{S}= \int dt\, \Big[ S_z\dot\phi + \Sb\!\int d^2r\, (\zeta\dot\xi - \xi\dot\zeta) \Big. - L_2\,(\delta S_z)^2 
- \tfrac12 \int d^2r\, (\xi\ \zeta)\, \mathcal{G}_{\Theta_0} \begin{pmatrix}\xi\\ \zeta\end{pmatrix}
- \delta S_z \int d^2r\, \mathcal{K}_{\Theta_0} \begin{pmatrix}\xi\\ \zeta\end{pmatrix}\big],\label{eq:collective_action}
\end{align}
where we expanded up to second order in terms of $\delta S_z=\mathcal{S}_z-\mathcal{S}_z^0$,$\mathcal{S}_z^0=\Sb\int d^2r (1-\cos\Theta_0)$. In the symmetric model of Eq.~\ref{eq:freeenergy}, the collective sector couples to the bath only through the helicity momentum $\delta S_z$, via the linear vertex $\mathcal{K}_{\Theta_0}$. Here $L_2$ is the coefficient for the quadratic part, and $\mathcal{G}_{\Theta_0}$ is the quadratic operator which determines the energy of the magnetic fluctuations. It is useful to introduce the complex magnon spinor,
\begin{equation}
\chi(\mathbf r,t)
=
\frac{1}{\sqrt{2}}
\begin{pmatrix}
\xi(\mathbf r,t)+i\zeta(\mathbf r,t) \\
\xi(\mathbf r,t)-i\zeta(\mathbf r,t)
\end{pmatrix}
\equiv U \mathbf{u}(\mathbf r,t),
\qquad
U=
\frac{1}{\sqrt{2}}
\begin{pmatrix}
1 & i \\
1 & -i
\end{pmatrix}.
\end{equation}
In this basis, the quadratic part of the free energy becomes $\frac{1}{2}
\mathbf{u}^{T}\mathcal{G}_{\Theta_0}\mathbf{u}
=\bar{S}\chi^\dagger \mathcal{H}_{\mathrm{mag}}\chi$
with $\mathcal{H}_{\mathrm{mag}}=\frac{1}{2\bar{S}} U \mathcal{G}_{\Theta_0} U^\dagger$. The magnon eigenmodes are obtained from the generalized pseudo-Hermitian paraunitary problem
\begin{equation}
\mathcal{H}_{\mathrm{mag}}\ket{\psi_n^{s}}
=\epsilon_n^s \sigma_z \ket{\psi_n^{s}}\,, \quad s=\pm1 \,,
\label{eq:magnon_eigenproblem}
\end{equation}
whose right/left eigenvectors obey the biorthogonality relation $\langle \psi^s_n|\sigma_z|\psi^{s'}_m\rangle = s\,\delta_{ss'}\delta_{nm}$ and has certain subtleties \cite{PhysRevB.100.134404,PhysRevX.7.041045}. Equation~\eqref{eq:magnon_eigenproblem} will allow us to evaluate the dissipative kernels as spectral sums over the magnon eigenmodes.

\section{Effective helicity dynamics from the Keldysh formalism} 
\label{sec:keldysh}

We now derive the effective real-time dynamics of the helicity after integrating out the magnon fluctuations. To this end, we formulate the action in Eq.~\eqref{eq:collective_action} on the closed Keldysh contour, with each dynamical field duplicated on the forward ($+$) and backward ($-$) branches \cite{Kamenev01052009}. We introduce the classical and quantum components
\begin{equation}
X^{\mathrm{cl}}=
\frac{X^{+}+X^{-}}{\sqrt{2}},
\qquad
X^{\mathrm{q}}=
\frac{X^{+}-X^{-}}{\sqrt{2}},
\label{eq:keldysh_rotation}
\end{equation}
for $X=\varphi,\delta S^{z},\chi$. In particular, the canonical term becomes
\begin{equation}
S^{z,+}\dot{\varphi}^{+}-S^{z,-}\dot{\varphi}^{-}=
\delta S^{z,\mathrm{q}}\dot{\varphi}^{\mathrm{cl}}+
\delta S^{z,\mathrm{cl}}\dot{\varphi}^{\mathrm{q}},
\end{equation}
where the contribution proportional to the equilibrium value $S_0^z$ is a boundary term and will be omitted. The magnon fields are collected into the Keldysh spinor $\Psi^T=
\begin{pmatrix} \chi^{\mathrm{cl}}, \chi^{\mathrm{q}} \end{pmatrix}$. Their quadratic action defines the usual retarded, advanced, and Keldysh Green functions,
\begin{equation}
\hat G=
\begin{pmatrix}
G^{K} & G^{R}\\
G^{A} & 0
\end{pmatrix},
\end{equation}
with $G^{R/A}(\omega)=\left[\mathcal{H}_{\mathrm{mag}}-(\omega\pm i0)\sigma_z\right]^{-1}$.
In thermal equilibrium, the Keldysh component is fixed by the fluctuation--dissipation relation
\begin{equation}
G^{K}(\omega)
=
\coth\left(\frac{\beta\omega}{2}\right)
\left[
G^{R}(\omega)-G^{A}(\omega)
\right].
\label{eq:FDT_G}
\end{equation}

In the $U(1)$-symmetric model, the helicity does not couple directly to the magnon field, and the only coupling between the collective and magnon sectors is mediated by its conjugate momentum $\delta S^z$. Thus, a Gaussian integration over $\Psi$ generates a nonlocal contribution to the collective-coordinate action, while the resulting effective action can be written as
\begin{align}
\mathcal{S}_{\mathrm{eff}}
={}&
\int dt\,
\left[
\delta S^{z,\mathrm{q}}\dot{\varphi}^{\mathrm{cl}}
+
\delta S^{z,\mathrm{cl}}\dot{\varphi}^{\mathrm{q}}
-
2L_2\,
\delta S^{z,\mathrm{cl}}
\delta S^{z,\mathrm{q}}
\right]
\nonumber\\
&+
\int dt\,dt'\,
\begin{pmatrix}
\delta S^{z,\mathrm{cl}}(t) &
\delta S^{z,\mathrm{q}}(t)
\end{pmatrix}
\begin{pmatrix}
0 & \Pi^{A}(t-t')\\
\Pi^{R}(t-t') & \Pi^{K}(t-t')
\end{pmatrix}
\begin{pmatrix}
\delta S^{z,\mathrm{cl}}(t')\\
\delta S^{z,\mathrm{q}}(t')
\end{pmatrix}.
\label{eq:effective_action_Sz}
\end{align}
Here the kernels $\Pi^{R,A,K}$ contain the complete influence of the magnon bath on the collective sector, and can be obtained in their spectral representation using the magnon eigenstates introduced in Eq.~\eqref{eq:magnon_eigenproblem}, 
\begin{equation}
\Pi^{R,A}(\omega)=-\frac{\bar S}{4}\sum_{s=\pm1}\sum_n's\,
\frac{|K_n^{s}|^2}{\omega-\varepsilon_n^{s}\pm i0},
\label{eq:PiR_spectral}
\end{equation}
where
$K_n^{s}=\int d^2r\,
\widetilde K_{\Theta_0}(\mathbf r)
\,\sigma_z\,
\psi_n^{s}(\mathbf r)$ and $\widetilde K_{\Theta_0}=K_{\Theta_0}U^\dagger$. The prime indicates that the collective zero mode, already accounted for explicitly through $\varphi$, is excluded from the sum. In thermal 
equilibrium it holds $\Pi^{K}(\omega)
=
\coth\left(\frac{\beta\omega}{2}\right)
\left[
\Pi^{R}(\omega)-\Pi^{A}(\omega)
\right]$. The frequency dependence of the collective-coordinate dynamics is entirely determined by the magnon spectrum and by the mode-dependent matrix elements $K_n^s$.

The Keldysh component of Eq.~\eqref{eq:effective_action_Sz} can equivalently be represented by a Gaussian stochastic field $\Xi(t)$ satisfying $\langle \Xi(t)\rangle=0$ and $
\langle \Xi(t)\Xi(t')\rangle=i\Pi^{K}(t-t')$. Variation of the effective action with respect to the quantum fields then gives the semiclassical equations of motion,
\begin{equation}
\Xi(t)=2L_2\,\delta S^{z,\mathrm{cl}}(t)-\dot{\varphi}^{\mathrm{cl}}(t)-2\int_{-\infty}^{t}dt'\,
\Pi^{R}(t-t')\delta S^{z,\mathrm{cl}}(t')\,,
\label{eq:langevin_Sz}
\end{equation}
The last term in Eq.~\ref{eq:langevin_Sz} is nonlocal in time and describes the memory induced by the magnon bath. Because $\delta S^{z,\mathrm{cl}}$ is conserved in the symmetric model, Eq.~\eqref{eq:langevin_Sz} reduces to
\begin{equation}
\Xi(t)
=
\left[
2L_2
-
\frac{\bar S}{2}
\sum_{s=\pm1}
\sum_n'
s\,
\frac{|K_n^{s}|^2}{\varepsilon_n^{s}}
\right]
\delta S^{z,\mathrm{cl}}
-
\dot{\varphi}^{\mathrm{cl}}(t).
\label{eq:langevin_spectral}
\end{equation}
Thus, in the $U(1)$-symmetric problem, the magnon bath modifies the relation between the helicity velocity and its conserved conjugate momentum, but does not generate a direct friction term proportional to $\dot{\varphi}$. In the following section, we integrate out $\delta S^z$ to formulate the dynamics solely in terms of the helicity and show explicitly that this correction corresponds to a renormalization of the helicity inertia.

\subsection{Magnon-induced mass renormalization}
\label{sec:mass_renormalization}

To isolate the dynamics of the helicity, it is convenient to integrate out its conjugate momentum $\delta S^z$ before eliminating the magnon fields. Since $\delta S^z$ enters quadratically in Eq.~\eqref{eq:collective_action}, this integration can be performed exactly. It generates a bare kinetic term for the helicity, a coupling between $\dot{\varphi}$ and the magnon fluctuations, and a correction to the quadratic magnon operator. After the magnon fields are integrated out, the effective Keldysh action for the
helicity takes the form
\begin{align}
S_{\mathrm{eff}}[\varphi]
={}m_0\int dt\,\dot{\varphi}^{\mathrm q}(t)\dot{\varphi}^{\mathrm{cl}}(t)
+\int dt\,dt'\,
\begin{pmatrix}
\dot{\varphi}^{\mathrm{cl}}(t) &
\dot{\varphi}^{\mathrm q}(t)
\end{pmatrix}
\begin{pmatrix}
0 &
\Pi_{\dot{\varphi}}^{A}(t-t')
\\
\Pi_{\dot{\varphi}}^{R}(t-t') &
\Pi_{\dot{\varphi}}^{K}(t-t')
\end{pmatrix}
\begin{pmatrix}
\dot{\varphi}^{\mathrm{cl}}(t')
\\
\dot{\varphi}^{\mathrm q}(t')
\end{pmatrix}.
\label{eq:Seff_phi}
\end{align}

The magnon contribution is nonlocal in time and is fully determined by the kernels $\Pi_{\dot{\varphi}}^{R,A,K}$. At the present order, the kernels in Eq.~\eqref{eq:Seff_phi} are obtained directly
from those of Eq.~\eqref{eq:effective_action_Sz} by rescaling the collective--magnon vertex, $\Pi_{\dot{\varphi}}^{R,A,K}=\frac{1}{4L_2^2}\, \Pi^{R,A,K}$, where we ignore corrections of order $K_{\Theta_0}^4$. The Keldysh component is fixed in equilibrium by the fluctuation--dissipation relation.

Although the magnon kernels are time (frequency) dependent, we are usually interested in the long-time asymptotic behavior of the helicity dynamics, which is in turn determined by the low-frequency part of the kernel, at frequencies well below the magnon gap, $|\omega|\ll\epsilon_{\mathrm{gap}}$. In this regime, the retarded and advanced kernels are analytic around $\omega=0$, and the effective equation of motion is $-m_{\mathrm{eff}}\,\omega^2 \varphi^{\mathrm{cl}}(\omega)+\mathcal{O}(\omega^3)=0$, where we introduced the renormalized helicity mass,
\begin{equation}
m_{\mathrm{eff}}=m_0+\Pi_{\dot{\varphi}}^{R}(0)+\Pi_{\dot{\varphi}}^{A}(0)=\frac{1}{2L_2}
+\frac{\bar S}{8L_2^2}\sum_{s=\pm1}\sum_n's\,\frac{|K_n^s|^2}{\epsilon_n^s}
+\mathcal O(K_{\Theta_0}^4).
\label{eq:meff_kernel}
\end{equation}

Equation~\eqref{eq:meff_kernel} expresses the magnon-induced correction entirely in terms of the eigenvectors of the unperturbed magnon Hamiltonian. Each mode contributes according to its overlap $K_n^s$ with the helicity-momentum vertex and inversely with its excitation energy. Thus, virtual magnon excitations dress the helicity dynamics and renormalize its inertia. Importantly, in the $U(1)$-symmetric model the bath couples to the collective sector only through $S^z$, and hence, after eliminating $S^z$, through $\dot{\varphi}$. The corresponding contribution to the inverse helicity propagator therefore starts at order $\omega^2$, rather than at order $\omega$. Moreover, within the undamped quadratic magnon theory, $\operatorname{Im}\Pi_{\dot{\varphi}}^R(\omega)=0$ below the magnon threshold. The magnon bath is consequently reactive in the low-frequency regime and does not generate Ohmic friction for the helicity.

\section{Ohmic channels due to $U(1)$-breaking terms}
\label{sec:U1breaking}

Real materials deviate from the ideal centrosymmetric limit through
Dzyaloshinskii--Moriya interactions (DMI), dipolar interactions, in-plane
anisotropies, or inhomogeneous fields. We now consider a generic static
perturbation that breaks the global $U(1)$ symmetry and therefore lifts the
conservation of $S^z$. We expand around a stationary helicity $\varphi_0$ by writing $\varphi=\varphi_0+\delta\varphi$. Taking the magnetic texture about which the fluctuations are defined to be the relaxed symmetry-broken saddle,
the terms linear in $\delta S^z$, $u$ and $\delta \varphi$ vanish at the stationary point. To quadratic order in the dynamical fluctuations, the action can be written as
\begin{align}
S_{\rm sb}=-\int dt \bigg[&V(\varphi)+L_{\rm sb}(\delta S^z)^2+A\,\delta S^z\delta\varphi+\frac{1}{2}\int d^2r\,u^T\Gamma u\nonumber\\
&+\delta S^z\int d^2r\,\mathcal{K}_{{\rm sb}}(\mathbf r)u(\mathbf r)+
\delta\varphi\int d^2r\,C_\varphi(\mathbf r)u(\mathbf r)\bigg],
\label{eq:Ssb_general}
\end{align}
where $C_\varphi(\mathbf r)$ is the direct helicity--magnon vertex,
\begin{equation}
C_\varphi(\mathbf r)=\left.\partial_\varphi C(\varphi,\mathbf r)\right|_{\varphi_0}=i\sum_{p\in\mathbb Z}p\,e^{ip\varphi_0}C_p(\mathbf r)=i\sum_{p\in\mathbb Z}p\,e^{ip\varphi_0}\sum_{\ell\in\mathbb Z}
C_{p\ell}(\rho)e^{i\ell\phi}.
\label{eq:Cphi}
\end{equation}
Equation~\eqref{eq:Cphi} makes explicit that $C_\varphi$ is a coherent sum over all helicity harmonics of the underlying symmetry-breaking interaction. Here $V(\varphi)$ is the helicity potential generated by the symmetry-breaking interaction fixing the helicity at $\varphi_0$, $\Gamma$ modifies the magnon Hamiltonian, and $\mathcal{K}_{\rm sb}$ modifies the momentum--magnon vertex. 

After integrating out the magnons, the nonlocal part of the effective helicity action is conveniently expressed in terms of
$\eta^{\rm cl/q}=(\delta\varphi^{\rm cl/q}, \dot{\varphi}^{\rm cl/q})^T$, with a retarded kernel of the form,
\begin{equation}
B^R(\omega)=\begin{pmatrix}
\Pi_{\varphi\varphi}^R(\omega) &
\Pi_{\varphi t}^R(\omega)\\
\Pi_{t\varphi}^R(\omega) &
\Pi_{tt}^R(\omega)
\end{pmatrix},
\end{equation}
and a spectral representation
\begin{equation}
\Pi_{\alpha\beta}^R(\omega)=-\frac{\bar S}{4}\sum_{s=\pm1}\sum_n's\,
\frac{V_{\alpha,n}^sV_{\beta,n}^{s*}}{\omega-\bar\epsilon_n^s+i0}.
\label{eq:Pi_ab_sb}
\end{equation}

Here, the projections of the collective vertices onto these modes are
\begin{equation}
V_{\alpha,n}^s=\int d^2r\,\widetilde V_\alpha(\mathbf r)\,\sigma_z\,\bar\psi_n^s(\mathbf r),\qquad\alpha=\varphi,t.
\label{eq:Valpha_modes}
\end{equation}
In particular, the direct helicity vertex contains the full Fourier sum
\begin{equation}
V_{\varphi,n}^s=i\sum_{p\in\mathbb Z}p\,e^{ip\varphi_0}C_{p,n}^s-\frac{A}{2L}\mathcal{K}_{{\rm tot},n}^s,
\label{eq:Vphi_harmonics}
\end{equation}
where $C_{p,n}^s=\int d^2r\,\widetilde C_p(\mathbf r)\,\sigma_z\,\bar\psi_n^s(\mathbf r)$ and $K_{{\rm tot},n}^s=\int d^2r\,\widetilde K_{\rm tot}(\mathbf r)\,\sigma_z\,\bar\psi_n^s(\mathbf r)$. It also holds $\widetilde V_t =\widetilde{\mathcal{K}}_{\rm tot}/2L$, $\widetilde V_\varphi=\widetilde C-(A/2L)\widetilde{\mathcal{K}}_{\rm tot}$, $\tilde{\mathcal{K}}_{\rm tot}=\tilde{\mathcal{K}}_{\rm \Theta_0}+\tilde{\mathcal{K}}_{\rm sb}$, $L=L_2+L_{\rm sb}$, and $\widetilde{\mathcal{K}}_{\rm tot}=
\mathcal{K}_{\rm tot}U^\dagger/\bar{S}$, $\widetilde{C}=CU^\dagger/\bar{S}$. The symmetry-breaking contribution $\Gamma$ modifies the magnon operator according to
$\mathcal H_{\rm mag}^{\rm sb}=\mathcal H_{\rm mag}+(1/2\bar{S})U\Gamma U^\dagger$ and we denote its eigenmodes by $\mathcal H_{\rm mag}^{\rm sb}\ket{\bar\psi_n^s}=\bar\epsilon_n^s\sigma_z
\ket{\bar\psi_n^s}$. The advanced and Keldysh components follow from $B^A=(B^R)^\dagger$ and the fluctuation--dissipation relation, respectively. Equation~\eqref{eq:Vphi_harmonics} show explicitly that the response kernels contain coherent contributions from all helicity harmonics and all magnon angular
sectors allowed by the spatial structure of the corresponding vertices.

Using $\delta\dot\varphi(\omega)=-i\omega\delta\varphi(\omega)$,
the magnon contribution to the retarded inverse helicity response is
\begin{equation}
\Sigma_\varphi^R(\omega)
=-2\bigg\{\Pi_{\varphi\varphi}^R(\omega)+i\omega\left[\Pi_{t\varphi}^R(\omega)
-\Pi_{\varphi t}^R(\omega)\right]+\omega^2\Pi_{tt}^R(\omega)\bigg\}.
\label{eq:Sigma_phi_general}
\end{equation}
The factor of two follows from the Keldysh convention used in
Eq.~\eqref{eq:effective_action_Sz}, where the retarded and advanced off-diagonal contributions are both retained explicitly. The corresponding Ohmic coefficient is defined by the low-frequency imaginary part,
\begin{equation}
\eta_{\rm Ohm}=-\lim_{\omega\rightarrow0}\frac{\operatorname{Im}\Sigma_\varphi^R(\omega)}{\omega}=2\left.\operatorname{Im}\partial_\omega\Pi_{\varphi\varphi}^R(\omega)\right|_{\omega=0}\,,
\end{equation}

where we used that in equilibrium $\operatorname{Re}[\Pi_{t\varphi}(0)]=\operatorname{Re}[\Pi_{\varphi t}(0)]$. 

Ohmic friction of the skyrmion helicity is a symmetry-breaking effect at its root, not merely a quantitative correction. The Langevin equation for the helicity then takes the familiar form
\begin{align}
m'\ddot\phi + \eta_{\rm Ohm}\dot\phi + \frac{\partial V}{\partial \phi} = \Xi(t),
\qquad
\langle \Xi(t)\Xi(t')\rangle = i\,\Sigma^K(t-t'),
\label{eq:langevin}
\end{align}
with colored noise correlations dictated by the fluctuation--dissipation theorem and the magnon density of states.

Thus, breaking $U(1)$ removes the symmetry protection against an Ohmic channel by allowing a direct helicity--magnon vertex. It does not, however, by itself guarantee a finite low-frequency friction coefficient. A static $U(1)$-breaking perturbation makes the direct helicity vertex $V_{\varphi,n}^{s}$ finite, but does not by itself provide a mechanism for irreversible energy transfer. For the conservative quadratic magnon problem and for a gapped magnon spectrum, it holds $\operatorname{Im}\Pi_{\varphi\varphi}^{R}(\omega)=0$, in a finite interval around $\omega=0$. The role of $U(1)$ breaking is therefore to remove the symmetry constraint that forbids a direct helicity--magnon coupling, rather than to generate dissipation by itself. A finite Ohmic coefficient additionally requires low-frequency dissipative spectral weight in the magnon bath. In the
following, we show that such spectral weight arises when the magnon modes have a finite lifetime.

\subsection{Effect of a finite magnon lifetime}
\label{sec:magnon_lifetime}

The discussion above assumes infinitely long-lived magnon eigenmodes. In a
realistic system, magnon--magnon, magnon--phonon, and other relaxation
processes give the magnon resonances a finite linewidth. We account for this
phenomenologically by assigning each mode a linewidth $\Gamma_n$, $\Gamma_n=(\bar{\epsilon}_n\tau_n)^{-1}$, where $\tau_n$ is the decay time of the magnon amplitude. In the eigenbasis of $\mathcal H_{\rm mag}^{\rm sb}$, the retarded and advanced magnon propagators therefore contain the poles $[\omega-\bar\epsilon_n^s+i\,s\Gamma_n\omega]^{-1}$, where we used a physical Gilbert-type broadening with frequency-dependent damping and a magnon resonance width at $\Gamma_n\bar{\epsilon}_n$.

The Ohmic coefficient is equal to 
\begin{equation}
\eta_{\rm Ohm}
=
\frac{\bar S}{2}
\sum_{s=\pm1}\sum_n'
\Gamma_n\,
\frac{\left|V_{\varphi,n}^{s}\right|^2}
{\left(\bar\epsilon_n^{s}\right)^2}
=
\bar S\sum_{n>0}'
\Gamma_n\,
\frac{\left|V_{\varphi,n}\right|^2}
{\bar\epsilon_n^{\,2}},
\label{eq:eta_Gamma}
\end{equation}

Thus the Ohmic contribution is linear in $\Gamma_n$ and vanishes continuously as $\Gamma_n\rightarrow0$. The role of $U(1)$ breaking and that of magnon relaxation are therefore distinct: symmetry breaking generates a direct
helicity--magnon vertex $V_{\varphi,n}^s$, whereas the finite linewidth
provides the low-frequency dissipative spectral weight required for a finite
Ohmic coefficient.

In the $U(1)$-symmetric limit, $V_{\varphi,n}^s=0$, irrespective of the finite lifetime. Magnon broadening can still generate an imaginary part of the velocity--velocity kernel $\Pi_{tt}^R$, but its contribution to the inverse helicity response is multiplied by $\omega^2$ and therefore remains super-Ohmic at low frequency.

\section{Selection rules and the superlattice spectrum}
\label{sec:spectrum}

The spectral representation of the mass correction also makes the symmetry constraints on the helicity--magnon coupling transparent. Since each magnon mode contributes only through its overlap with the vertex $\widetilde{\mathcal{ K}}$ or $V_{\varphi,n}^s$, the spatial symmetry of this vertex determines which sectors of the excitation spectrum can enter the dissipative terms. This provides a direct selection rule relating the collective helicity dynamics to specific magnon modes.

\subsection{Angular selection rule and the breathing channel}
  
For a rotationally symmetric skyrmion profile $\Theta_0(\rho)$ in the continuum theory, $\mathcal{K}_{\Theta_0}$ is radial and therefore carries zero angular momentum. Consequently, the matrix elements $K_n^s$ entering Eq.~\eqref{eq:meff_kernel} select the $\ell=0$ sector of the magnon spectrum,
corresponding to breathing-symmetry fluctuations. Modes with
$\ell\neq0$, including the gyration modes, do not contribute in the
continuum rotationally invariant limit. On the microscopic triangular
lattice this selection rule is reduced to the corresponding discrete
rotational symmetry.

The nature of the $\ell=0$ sector differs substantially between chiral
and frustrated magnets. In DMI-stabilized skyrmions the breathing mode
appears as a localized magnon bound state below the continuum \cite{Garst_2017}.
The nature of the breathing-symmetry sector differs substantially between
chiral and frustrated magnets. In the frustrated system considered in
Ref.~\cite{Hullahalli2026}, the Mexican-hat dispersion possesses six
symmetry-related minima at finite wave vector $k_{\rm min}$, with a
characteristic wavelength $l_{\rm min}=2\pi/k_{\rm min}$ comparable to the
skyrmion size. As a consequence, skyrmion deformation modes are pushed into
the magnon continuum and hybridize with extended finite-$k$ states. The lowest breathing-type excitation lies at $\omega_{\rm BM}\approx\omega_{\rm gap}$ and is not an isolated bound state but a skyrmion-hybridized resonance embedded in the continuum, dressed by the interference pattern of the finite-$k_{\rm min}$ states. 

This structure has two consequences for the mass kernel. First, matching the skyrmion and magnon length scales, $l_{\rm min}\sim\lambda$, can enhance the overlap between the breathing-sector eigenmodes and the radial vertex $\mathcal{K}_{\Theta_0}(\mathbf r)$. Second, the relevant modes occur close to the magnon threshold, $\omega_{\rm gap}=\bar{B}_z+2\bar{K}-\bar{J}_2$, so that the inverse-energy weighting in Eq.~\eqref{eq:meff_kernel} favors them. Reducing the gap toward the critical field can therefore enhance the helicity mass renormalization, providing an experimentally tunable knob for the helicity inertia. These considerations suggest that the magnon-induced helicity mass in
frustrated magnets is controlled predominantly by breathing-symmetry
modes hybridized with the finite-$k$ continuum.

\subsection{Selection rules for $U(1)$-symmetry-breaking terms}

The harmonic decomposition of Eq.~\eqref{eq:Cphi} makes it possible to identify
which sectors of the magnon spectrum enter the Ohmic coefficient for a
given $U(1)$-breaking perturbation. The helicity harmonic $p$
and the spatial angular harmonic $\ell$ are independent. We illustrate this
distinction with three representative perturbations.

\paragraph{Dzyaloshinskii--Moriya interaction.}
Let's consider first an isotropic bulk Dzyaloshinskii--Moriya interaction,
$F_{\rm DMI}=D\int d^2r\,{\bf m}\cdot(\nabla\times{\bf m})$. For the axisymmetric $Q=-1$ texture, $\Phi=\phi+\varphi$, the DMI
depends on the helicity through the first harmonics,
$p=\pm1$. After expanding to linear order in the magnon fluctuations,
however, the explicit dependence on the spatial polar angle cancels
against the winding of the background texture. At the DMI-selected
equilibrium helicity the direct vertex takes the radial form
\begin{equation}
C_D^{\rm }(\rho)=-2\sin\varphi_0 \sin\Theta_0(\rho)\Theta_0'(\rho)(1,0),
\end{equation}
Thus, $p=\pm1$ and $\ell=0$. Equation~\eqref{eq:eta_Gamma} therefore gives
\begin{equation}
\eta_{\rm Ohm}^{\rm DMI}=\frac{\bar S D^2}{2}\sum_{s=\pm1}\sum_{n,\ell=0}'
\Gamma_n\frac{\left|C_{D,n}^{s}\right|^2}{\left(\epsilon_n^{s}\right)^2}
+\mathcal{O}(D^3).
\label{eq:eta_DMI_direct}
\end{equation}
where we used the mode-resolved matrix element ${\cal C}_{D,n}^{s}=\int d^2r\,
\widetilde{\cal C}_{D}(\rho)\sigma_z\psi_n^s(\mathbf r)$. The radial structure of the direct helicity vertex restricts this contribution to the $\ell=0$ sector.

\paragraph{In-plane anisotropy.}
Consider next a uniaxial in-plane anisotropy, $F_{K_x}=-K_x\int d^2r\,m_x^2$. The helicity--magnon vertex contains second harmonics in both the helicity
and spatial angles,
\begin{equation}
C_\varphi^{K_x}(\rho,\phi)
=
K_x
\left(
\alpha(\rho)\sin^2\Theta_0
\cos[2(\phi+\varphi_0)],
\,
\beta(\rho)\sin\Theta_0\cos\Theta_0
\sin[2(\phi+\varphi_0)]
\right).
\end{equation}
Consequently, $p=\pm2$ and $\ell=\pm2$. The anisotropy itself both modifies the momentum vertex and mixes magnon states with angular momenta differing by $\Delta\ell=\pm2$. Their leading contribution to Eq.~\eqref{eq:eta_Gamma} is therefore,
\begin{align}
\eta_{\rm Ohm}^{K_x}=\frac{\bar S K_x^2}{2}\sum_{s=\pm1}
\sum_{n,|\ell|=2}'\Gamma_n\frac{\left|C_{K_x,n}^{s}\right|^2}{\left(\epsilon_n^{s}\right)^2}+\mathcal{O}(K_x^3).
\label{eq:eta_Kx}
\end{align}
where we used $V_{\varphi,n}^{s}=K_x{C}_{K_x,n}^{s}$ for modes originating from $|\ell|=2$. The leading weak-anisotropy scaling is consequently $\eta_{\rm Ohm}^{K_x}\propto K_x^2/\tau_{\rm mag}$. At finite $K_x$ the equilibrium skyrmion becomes elliptical, so that $\ell$ is no longer an exact quantum number. The perturbation preserves only the corresponding twofold symmetry and mixes sectors differing by $\Delta\ell=2$; the strict $|\ell|=2$ rule above should therefore be understood as the weak-$K_x$ limit of coupling to the even $C_2$ sector.

\paragraph{Static in-plane magnetic field.}
A direct route to the dipolar magnon sector is provided by a uniform
static in-plane field, $F_{h_x}=-h_x\int d^2r\,m_x$. The linearized
helicity--magnon vertex is
\begin{equation}
C_\varphi^{h_x}(\rho,\phi)=h_x\left(\alpha(\rho)\sin\Theta_0
\cos(\phi+\varphi_0),\,\beta(\rho)\cos\Theta_0\sin(\phi+\varphi_0)\right).
\end{equation}
The perturbation therefore carries $p=\pm1$ and $\ell=\pm1$. The direct vertex consequently couples the helicity to the dipolar sector containing the clockwise and counterclockwise gyration modes. Eq.~\eqref{eq:eta_Gamma} gives
\begin{align}
\eta_{\rm Ohm}^{h_x}=\frac{\bar S h_x^2}{2}
\sum_{s=\pm1}\sum_{n,|\ell|=1}'\Gamma_n\frac{\left|C_{h_x,n}^{s}\right|^2}{\left(\epsilon_n^{s}\right)^2}
+\mathcal{O}(h_x^3)
\label{eq:eta_hx}
\end{align}
where we used $V_{\varphi,n}^{s}=h_x{C}_{h_x,n}^{s}$. Hence a weak static in-plane field produces $\eta_{\rm Ohm}^{h_x}\propto h_x^2/\tau_{\rm mag}$. Because a real field contains both $\ell=+1$ and $\ell=-1$ components, it allows both CW and CCW modes by symmetry. Their relative contribution to Eq.~\eqref{eq:eta_hx} is determined by the mode-resolved matrix elements. In the frustrated-skyrmion spectrum, this provides a direct mechanism by which the magnetically active CCW modes can contribute to the helicity friction.

These examples demonstrate that $U(1)$ breaking alone does not determine
the magnon modes entering the Ohmic response. In the weak-perturbation
limit, bulk DMI has $(|p|,|\ell|)=(1,0)$, in-plane anisotropy has
$(2,2)$, and a uniform in-plane field has $(1,1)$. The spatial harmonic
$\ell$ fixes the leading magnon selection rule, while the helicity
harmonic $p$ determines how the corresponding vertex arises from the
periodic dependence on the collective coordinate. Together with
Eq.~\eqref{eq:eta_Gamma}, these selection rules determine both the modes and the
perturbative order at which each symmetry-breaking mechanism contributes
to $\eta_{\rm Ohm}$.

\section{Estimates and implications for helicity qubits}
\label{sec:estimates}

For the parameter set of Ref.~\cite{Hullahalli2026} ($\bar{J}_1=2$~meV, $\bar{J}_2=1.5$~meV, $\bar{K}=0.3$~meV, $a=0.5$~nm), the magnon gap is $\omega_{\rm gap}\simeq0.05$~meV ($\simeq12$~GHz) at $\bar{B}_z=3.45$~T and the skyrmion core diameter is
$\lambda\simeq6$~nm. The helicity-qubit level splittings proposed in
Refs.~\cite{PhysRevLett.127.067201,10.1063/5.0177864} lie in the low-GHz range,
below the one-magnon threshold. In the undamped quadratic theory this
implies $\operatorname{Im}\Pi^R(\omega)=0$ at the relevant low
frequencies, while for $\omega\ll\omega_{\rm gap}$ the local
low-frequency expansion applies.

In the $U(1)$-symmetric problem, the leading effect of the magnon bath is
the mass renormalization of Eq.~\eqref{eq:meff_kernel}. This is a reactive
rather than dissipative contribution: virtual magnon excitations modify
the effective helicity inertia and therefore shift the tunnelling
amplitude and transition frequencies. A finite magnon linewidth
does not change the symmetry restriction on the helicity vertex. A static $U(1)$-breaking perturbation opens additional helicity--magnon channels but, as shown in Sec.~\ref{sec:U1breaking}, does not by itself produce irreversible Ohmic damping. For a gapped quadratic magnon bath, $\eta_{\rm Ohm}=0$ in the limit
$\tau_n\rightarrow\infty$. Ohmic damping arises when the magnon modes have a finite lifetime, with its magnitude controlled by the
strength and symmetry of the $U(1)$-breaking perturbation, the
mode-resolved helicity--magnon overlaps, and the intrinsic magnon
lifetimes.


Upon quantization of the helicity coordinate, the Ohmic coefficient derived
here connects directly to the phenomenological decoherence theory of
Ref.~\cite{PhysRevLett.127.067201}. There, Gilbert damping was projected onto the
collective helicity dynamics and represented by an Ohmic noise spectrum,
leading within Bloch--Redfield theory to a relaxation rate of the form
\begin{equation}
T_1^{-1}
\propto
\alpha_\varphi
\left(\gamma_\varphi^\perp\right)^2
\omega_{01}
\coth\left(\frac{\beta\omega_{01}}{2}\right),
\end{equation}
where $\omega_{01}=(E_1-E_0)/\hbar$ is the transition frequency between
the two lowest qubit eigenstates and $\gamma_\varphi^\perp$ contains the
corresponding helicity transition matrix element. The coefficient
$\eta_{\rm Ohm}$ obtained here provides the microscopic magnon contribution
corresponding to the phenomenological helicity damping coefficient
$\alpha_\varphi$. The microsecond coherence times estimated in
Ref.~\cite{PhysRevLett.127.067201} for an ultralow Gilbert damping
$\alpha_{\rm G}\sim10^{-5}$ may therefore be regarded as phenomenological
benchmarks. The present theory instead relates the effective damping to the
symmetry-breaking helicity--magnon matrix elements and to the lifetimes of
the relevant magnon modes, allowing the corresponding qubit relaxation rate
to be evaluated microscopically.

The mode dependence of the kernels provides a direct spectroscopic
route to identifying the dominant dissipation channels. The relevant
probe depends on the spatial symmetry of the perturbation. For DMI,
which couples the helicity to the $\ell=0$ sector, the breathing
resonances are the natural modes to monitor. A static in-plane field,
on the other hand, couples to the $|\ell|=1$ sector and therefore makes
the magnetically active CW and CCW resonances directly relevant to the
helicity response. Measurements of the corresponding magnon resonance
frequencies, linewidths and spectral weights, together with the
helicity linewidth or ring-down time, can therefore constrain both the
mode-resolved couplings and the magnon lifetimes entering
$\eta_{\rm Ohm}$. 
 
\section{Conclusions}
\label{sec:conclusion}
 
We have developed a microscopic description of skyrmion-helicity dynamics
coupled to the magnon environment in centrosymmetric frustrated magnets.
Starting from the collective helicity coordinate and its conjugate spin
$S^z$, we integrated out the magnon fluctuations within the real-time
Keldysh formalism and expressed the resulting response kernels directly in
the magnon eigenbasis. In the $U(1)$-symmetric model, the coupling to the
bath occurs only through the helicity momentum and therefore becomes a
derivative coupling after $S^z$ is eliminated. To leading order, virtual
magnon excitations consequently generate a reactive renormalization of the
helicity mass, with each mode weighted by its overlap with the collective
vertex and inversely by its excitation energy. For an undamped gapped magnon
spectrum, the corresponding low-frequency response contains no Ohmic
friction.

We further showed that the effect of $U(1)$-breaking perturbations is most
naturally characterized by separating their harmonics in helicity space
from their spatial angular harmonics. A static symmetry-breaking term opens
a direct helicity--magnon vertex and mixed helicity--velocity response
kernels, but does not by itself generate irreversible damping: for a
conservative gapped quadratic magnon bath the Ohmic coefficient remains
zero. A finite magnon lifetime supplies the low-frequency spectral weight
required for a finite $\eta_{\rm Ohm}$, which in the weak-broadening regime
scales with the inverse magnon lifetime. The magnitude of the resulting
friction is therefore determined jointly by the symmetry-breaking
matrix elements, the magnon spectrum, and the magnon lifetimes, rather than
by the strength of $U(1)$ breaking alone.

The harmonic structure of the symmetry-breaking interaction additionally
provides mode-resolved selection rules for the dissipative kernels. In the
continuum limit, isotropic bulk DMI carries helicity harmonics $p=\pm1$ but
a radial spatial vertex, and therefore couples directly to the $\ell=0$
sector. A weak in-plane anisotropy $m_x^2$ carries $p=\pm2$ and
$\ell=\pm2$ at leading order around the circular skyrmion, while a uniform
in-plane field $m_x$ carries $p=\pm1$ and $\ell=\pm1$ and hence provides a
direct coupling to the dipolar sector containing the CW and CCW modes.
At finite symmetry breaking, the deformation of the equilibrium texture
mixes the corresponding angular sectors, but the weak-perturbation
harmonics remain useful for identifying the dominant magnon channels.
This establishes a direct connection between the spatial form of the
microscopic perturbation and the subset of the skyrmion excitation spectrum
that controls helicity dissipation.

These results provide a microscopic framework for assessing dissipation in
helicity-based skyrmion qubits. In the symmetric limit, the magnon bath
primarily renormalizes the helicity inertia, whereas Ohmic relaxation
requires both a symmetry-breaking coupling and dissipative magnon spectral
weight. The theory therefore identifies two complementary routes for
coherence engineering: suppressing perturbations that couple the helicity
to magnetically active magnon sectors and increasing the lifetime of the
relevant magnon modes. Conversely, controlled symmetry breaking provides a
way to selectively couple the helicity to particular excitations and may
be useful for spectroscopy, initialization, and relaxation control.
Combining mode-resolved measurements of magnon frequencies and linewidths
with helicity spectroscopy would provide a direct experimental test of the
microscopic kernels derived here. Extensions to interacting magnons,
phonon and electronic environments, and the quantization of the resulting
non-Markovian helicity dynamics provide natural directions for determining
$T_1$ and $T_2$ of realistic skyrmion-based quantum states.

\section{Acknowledgement}

C.P. is an \'{E}cole Normale Sup\'{e}rieure (ENS)-Mitsubishi Heavy Industries (MHI) Chair of Quantum Information supported by MHI.


\bibliographystyle{RS}
\bibliography{sample}
\end{document}